\documentclass[aps,pre,twocolumn,superscriptaddress,nofootinbib]{revtex4-1}
\usepackage{graphicx}
\usepackage{epstopdf}
\usepackage{soul}
\usepackage{color}
\usepackage{amsmath}
\usepackage{lineno}
\usepackage{dcolumn}
\usepackage{bm}
\usepackage{float}
\usepackage{xcolor}

\begin{document}

\title{Fluctuation-theorem and extended thermodynamics of turbulence}


\author{Amilcare Porporato}
\email{aporpora@princeton.edu}
\affiliation{Department of Civil and Environmental Engineering and Princeton Environmental Institute, Princeton University, Princeton, NJ 08544, USA}
\author{Milad Hooshyar}
\email{hooshyar@princeton.edu}
\affiliation{Princeton Environmental Institute and Princeton Institute for International and Regional Studies, Princeton University, Princeton, NJ 08544, USA}
\author{Andrew D Bragg}
\email{andrew.bragg@duke.edu}
\affiliation{Department of Civil and Environmental Engineering, Duke University, Durham, NC 27708, USA}
\author{Gabriel Katul}
\email{gaby@duke.edu}
\affiliation{Department of Civil and Environmental Engineering and Nicholas School of the Environment, Duke University, Durham, NC 27708, USA}




\begin{abstract}
Turbulent flows are out-of-equilibrium because the energy supply at large scales and its dissipation by viscosity at small scales create a net transfer of energy among all scales. Here, the energy cascade is approximated by a combined contribution of a forward drift and diffusion that recover accepted phenomenological theories of turbulence. The fluctuation theorem (FT) is then shown to describe the scale-wise statistics of forward and backward energy transfer and their connection to irreversibility and entropy production. The ensuing turbulence entropy may be used to formulate an extended turbulence thermodynamics.
\end{abstract}


\maketitle
\section{Introduction}
It is perhaps not coincidental that one of the most influential experiments in the history of thermodynamics is also a turbulence experiment. In 1849, James Prescott Joule used a stirrer to show that the shaft work on a fluid ends up increasing its internal energy, thereby demonstrating the equivalence of heat and work. Dealing with the generation of turbulent kinetic energy ($K$) and its subsequent dissipation rate ($\epsilon$) by viscosity, the Joule experiment also offers a modern link between thermodynamics, a theory of the macroscopic effects of microscopic fluctuations, and non-equilibrium fluctuations, of which turbulence is a quintessential example.  On the one hand, the disparity between microscopic and macroscopic fluctuations appears un-reconcilable. Thermodynamic fluctuations are so small to allow a mathematical description of fluids as a continuum. For this reason, turbulence is conveniently described by the Navier-Stokes equations assuming local thermodynamic equilibrium \citep{panton2006incompressible}. Turbulent fluctuations are thus of macroscopic nature and technically outside the scope of traditional thermodynamics \citep{sreenivasan1999fluid}. On the other hand, the random-like nature of turbulence \citep{Batchelor1953,monin1975statistical,Hinze_1959} invites a thermodynamic formalism to the problem of turbulence, including the eddy thermodynamics of Richardson \citep{richardson1920supply}, Blackadar \citep{blackadar1955extension} and others \citep{nevzglyadov1961theory,reinke2018universal}, as well as the Onsager analysis of 2D turbulence \citep{eyink2006onsager}. \citet{liepmann1979rise} asserted that `turbulence can be defined by a statement of impotence reminiscent of the second law of thermodynamics'. More recently, a number of studies have argued that the fluctuation theorem derived for small systems \citep{evans2002fluctuation,gallavotti1995dynamical} can be partly applied to describe macroscopic fluctuations so as to explore their time reversibility at multiple scales, including turbulence.

The statistical properties of turbulence differ from systems near thermal equilibrium because the flux of energy per unit mass is supplied at scales much larger than the scales at which energy is dissipated by the action of viscosity, resulting in an energy flux (cascade) across all scales. Such a transport is linked to multiple processes, including vortex stretching, self-amplification of the strain-rate and viscous diffusion \citep{tennekes1972first,carbone2020vortex}. One of the defining features of the turbulence cascade is that the probability of forward and backward transitions between two energetic states at a given scale are not identical (i.e., a scale-wise `detailed balance' is not applicable \citep{xu2014flight,fuchssmall}).

The objective of this work is to illustrate that non-equilibrium thermodynamics, and in particular, the fluctuation theorem can be extended to describe the behavior of the turbulent energy cascade.  While the net transfer of energy from large to small scales is prevalent, it is shown here that back-scatter of energy and its connection to time-scale irreversibility obeys the statistics predicted by the fluctuation theorem \citep{evans1993probability,gallavotti1995dynamical, porporato2007irreversibility,zorzetto2018extremes,fuchssmall}. To provide a physical context, a turbulent flow conceptually analogous to Joule's original experiment is used, where work is done on the fluid system to generate $K$ in a narrow band of scales, which is then dissipated as heat thereby raising the internal energy. In the analysis here, constant energy is externally supplied at a pre-selected scale much larger than the Kolmogorov length scale ($\eta_K$) where viscous effects are significant.  At steady state, the energy cascade develops in a manner where the energy injection rate is balanced by viscous dissipation rate as in Joule's experiment.  The fluctuation theorem is then applied to describe the forward and backward probabilities of energy packets moving scale-wise in time through the energy cascade. For analytical tractability and to illustrate connections with the fluctuation theorem, simplified closure schemes for the energy transfer rate across scales are employed. These closure schemes offer plausible expressions for the energy cascade that are consistent with a wide range of experiments and theories on locally homogeneous and isotropic turbulence.

\section{Spectral Energy Balance}

For a homogeneous, isotropic turbulent flow of a Newtonian, incompressible viscous fluid, the spectral energy balance per unit mass of fluid is \citep{Batchelor1953,Panchev1971,monin1975statistical}
\begin{equation}\label{eq:spectral}
\frac{\partial E(k)}{\partial t}=p(k)+\vartheta(k)-\eta(k),
\end{equation}
where $E(k)$ is the turbulent kinetic energy per unit wave number $k$, $p(k)$ is the production spectrum here assumed to be concentrated at $k=k_i$, $\vartheta(k)$ is the energy transfer spectrum, $\eta(k)$ is the viscous dissipation spectrum, and $k$ is the wavenumber or inverse eddy-size.
The normalizing property
$\int_0^{\infty} E(k) dk=({1}/{2})K$ defines the turbulent kinetic energy $K$.  Eq. (\ref{eq:spectral}) makes no other assumptions about the velocity statistics other than homogeneity and isotropy.  Because $\vartheta$ is a scale-wise transport and cannot contribute to production or destruction of $K$, it satisfies the integral constraint
$\int_0^{\infty} \vartheta dk=0$. It can be expressed as the gradient of an energy flux $J$,
\begin{equation}\label{eq:enflux}
\vartheta(k)=-\frac{\partial J}{\partial k},
\end{equation}
while the scale-wise viscous dissipation rate is given by
\begin{equation}\label{eq:dissk}
\eta(k)=2\nu k^2 E(k).
\end{equation}
 
Integrating Eq. \eqref{eq:spectral} over $k$ yields the energy balance of $K$,
\begin{equation}\label{eq:cont1}
\frac{d K}{d t}=w-\epsilon,
\end{equation}
where $w=\int_0^{\infty}{p(k)}dk$ is the rate of work done on the fluid to produce turbulence and $\epsilon$ is the dissipation rate of $K$,
\begin{equation}\label{eq:dissk1}
\epsilon=\int_0^{\infty}{\eta(k)}dk.
\end{equation}
The concomitant balance for internal energy $U$ is then $dU/dt=\epsilon -q$, where $q$ is the heat loss to the environment.  Because of fluid incompressibility, temperature fluctuations resulting from dissipation have no feedback on the dynamics of the turbulence, including the energy cascade.  Finally, the entropy balance is given by ${dS}/{dt}=-{q}/{T} +\sigma$,
where $T$ is the absolute temperature and $\sigma={\epsilon}/{T}$ is the entropy production.
It is assumed that $q$ is immediately delivered to a surrounding environment, acting as a thermal bath at the same temperature, thus ensuring isothermal conditions. 

Returning to the spectral energy balance, a closure of minimal complexity that preserves both direct energy cascade and an inverse cascade (or backscatter) may be obtained by representing the contributions to $J$ as scale-wise drift and a diffusion term linked by a timescale of eddy relaxation $\tau(k)$.  A flexible form for such a closure is proposed here as
\begin{equation}\label{eq:cont}
J=\alpha \frac{kE(k)}{\tau(k)}-\frac{k^2}{\tau(k)}\frac{\partial E(k)}{\partial k},
\end{equation}
where the coefficient $\alpha$ is to be determined depending on models for $\tau(k)$ \citep{Onsager1949,Corrsin1964,Antonia2009,poggi2011role}.  Substituting $J$ and $\eta$ from Eqs. \eqref{eq:cont} and \eqref{eq:dissk} into the spectral energy balance in Eq. \eqref{eq:spectral} yields
\begin{multline}
\frac{\partial E}{\partial t} =\left[p-2\nu k^2 E(k)\right]-\\ \frac{\partial }{\partial k}\left(\alpha \frac{kE(k)}{\tau(k)}-\frac{k^2}{\tau(k)}\frac{\partial E(k)}{\partial k}\right).
\label{eq:budg}
\end{multline}
Depending on the choices made about $\tau(k)$, a general class of non-linear diffusion models for $J(k)$ can be recovered. Here, a $\tau(k)$ that is linked to $E(k)$ is adopted,
\begin{equation}\label{eq:tau_leith}
\tau(k)=\left[k^3 E(k)\right]^{-1/2}.
\end{equation}
For the inertial subrange scales, the Kolmogorov  \citep{kolmogorov41} scaling (hereafter referred to as K41 scaling) given by $E(k)=C_o \epsilon^{2/3} k^{-5/3}$  is expected to hold resulting in $\tau(k)={C_o}^{-1/2} \epsilon^{-1/3} k^{-2/3}$ (i.e. Onsager's relaxation time \citep{eyink2006onsager}), where $C_o=1.55$ is the Kolmogorov constant. Due to the dissipative anomaly \citep{Falkovich2001}, $\lim_{\nu\to 0}\epsilon$ is finite, so that in this limit $\eta_K=(\nu^3/\epsilon)^{1/4} \rightarrow 0$, $(\eta_K)^{-1}\rightarrow\infty$. We may then estimate the total time for energy to be passed from a finite $k_i$ to an infinitely high wavenumber by 
\begin{equation}\label{eq:tau_leith2}
\int_{1/k_i}^{\infty} \tau(k)\frac{dk}{k} = \frac{3}{2}\frac{1}{\sqrt{C_o}} k_i^{2/3} \epsilon^{1/3} < \infty.
\end{equation}
Eq. (\ref{eq:tau_leith2}) implies that the steps in the energy cascade rapidly accelerate such that (if not interrupted by the action of viscosity at a finite wavenumber) the time for energy to be passed to an infinitely high wavenumber is finite.  This finding, originally put forth by Onsager \citep{eyink2006onsager}, foreshadows the finite time singularity in the inviscid limit for such classes of $\tau(k)$ models \citep{davidson2015turbulence}. The $\tau(k)$ in (\ref{eq:tau_leith}) is also singled out because it recovers the well-studied Leith's non-linear diffusion approximation \citep{Leith1968,besnard1996spectral,clark2009reassessment}, 
\begin{multline}\label{'eq:bal_leith_1'}
\frac{\partial}{\partial k}\left[k^{11/2} \sqrt{E(k)} \left(\alpha \frac{E(k)}{k^3}-\frac{1}{k^2}\frac{\partial E(k)}{\partial k}\right) \right]=\\ \frac{\partial E}{\partial t} - \left[p-2\nu k^2 E(k)\right].
\end{multline}
When $\alpha=2$, the conventional form of Leith's model becomes evident \citep{Leith1968,Panchev1971}. The latter recovers the so-called warm cascade condition (i.e. a steady equipartitioned energy spectrum, $\forall k: E(k)\propto k^2$) originally derived by Lee \citep{lee1952some} under specific conditions \citep{besnard1996spectral,connaughton2004warm,clark2009reassessment}. Leith's model was also derived from the so-called direct-interaction approximation when a number of simplifications are made \citep{clark2009reassessment}. 

For stationary conditions, and far from the production and viscous subranges, ${\partial E}/{\partial t} - p-2\nu k^2 E(k)=0$, the solution to the spectral budget reduces to
\begin{equation}\label{eq:sol_leith_sim}
E(k) = \left(C_1 k^{-\frac{5}{2}}+ C_2 k^{\frac{3}{2}\alpha}\right)^{\frac{2}{3}}.
\end{equation}
If $p$ is injected at $k=k_i$, then for $k>k_i$ $C_1=C_o^{3/2} \epsilon$ necessitating $C_2=0$ to recover K41 inertial subrange scaling (also referred to as the cold cascade). For $k<k_i$, $C_1=0$ and $C_2=C_o \epsilon^{2/3} k_i^{-11/3}$ is set by the continuity of $E(k)$ at $k_i$ to achieve a warm cascade for $\alpha = 2$ \citep{connaughton2004warm}. The $E(k)\sim k^{+2}$ is also compatible with the well-known Saffman spectrum \citep{saffman1967large,meyers2008functional,davidson2015turbulence}, a scaling law derived from considerations (continuity and smoothness) of how $E(k)$ is approached as $k \rightarrow 0$. 

\begin{figure}
\begin{centering}
\includegraphics[scale=0.5]{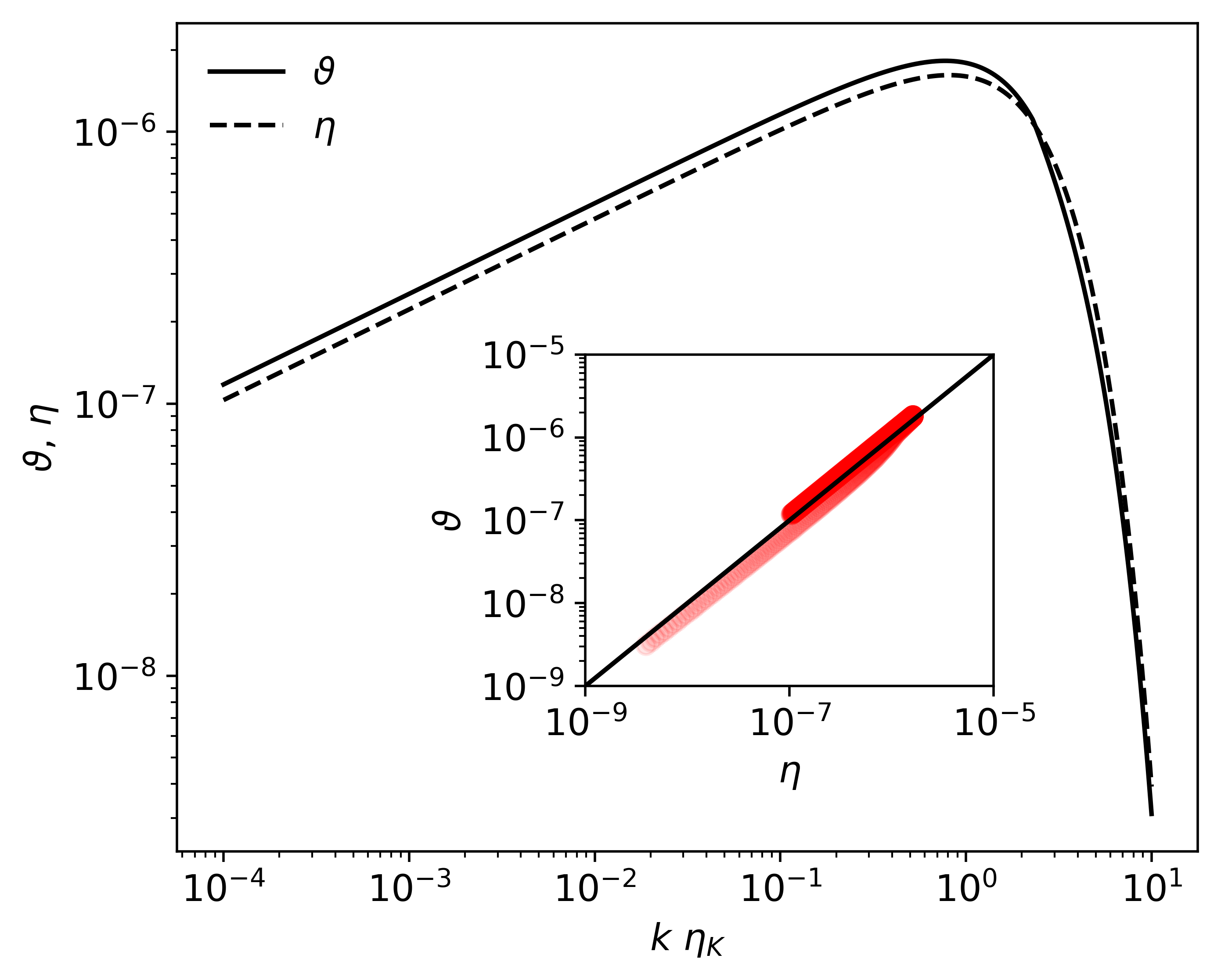}
\caption{The balance between energy transfer $\vartheta=-{dJ}/{dk}$ and viscous dissipation $\eta(k)$ across scale based the empirical spectrum in Eq. \eqref{eq:Sol_emp} for $k_i \eta_K = 10^{-4}$. Here, $\beta=$ 0.33 results in an acceptable spectral energy balance closure at steady state. The numerical value of $\beta$ here differs from the original Pao constant because of the choices made when deriving $\tau(k)$. The inset shows the one-to-one correlation between $\vartheta(k)$ and $\eta(k)$. }
\label{fig:emprical_sol}
\end{centering}
\end{figure}

In the presence of viscous dissipation, the spectral budget equation is not analytically solvable, however, we numerically confirm that 
\begin{equation}\label{eq:Sol_emp}
E_*(k) \approx \left(C_o^{3/2} \epsilon  k^{-\frac{5}{2}}+ C_2 k^{3}\right)^{\frac{2}{3}} f_{\eta}(k\eta_K)
\end{equation}
reasonably approximates the spectral energy budget, as shown in Fig \ref{fig:emprical_sol}.  Here, $f_{\eta}(k \eta_K)=\exp\left[{-\beta (k \eta_K)^{4/3}}\right]$ is the Pao correction \citep{Pao1965} reshaping the $k^{-\frac{5}{3}}$ spectrum for $k\eta_K>0.1$ \citep{Pope2005}. The $E_*(k)$ from Eq. \eqref{eq:Sol_emp} implies that $\tau(k)$ in Eq. \eqref{eq:tau_leith} increases within the viscous sub-range when $k \eta_K>\beta^{-3/4}$, which is not physically plausible. The increase in $\tau(k)$ is expected when $E_*(k)$ decreases faster than $k^{-3}$ with increasing $k$. Hence, an amendment proposed by Batchelor \citep{Batchelor1959} was used in the calculations featured in Fig \ref{fig:emprical_sol} whereby the straining rate ($\propto \tau(k)^{-1}$) at $k$ is assumed to be uniform beyond scales commensurate with $1/\eta_K$. This amendment revises the model for $\tau(k)$ as 
\begin{equation}\label{eq:emp_tau}
  \tau_*(k) =
  \begin{cases}
         C_2^{-\frac{1}{2}} k^{-\frac{5}{2}} f_{\eta}(k \eta_K)^{-\frac{1}{2}} &  k \eta_K < k_i \eta_K \\
        C_o^{-\frac{1}{2}}\epsilon^{-\frac{1}{3}} k^{-\frac{2}{3}} f_{\eta}(k \eta_K)^{-\frac{1}{2}} & k_i \eta_K < k \eta_K\leq\beta^{-\frac{3}{4}} \\
        \sqrt{\frac{e \beta}{C_o}} \tau_K & k \eta_K>\beta^{-\frac{3}{4}},
  \end{cases}
\end{equation}
where $\tau_K=(\nu/\epsilon)^{1/2}$ is the Kolmogorov time scale. 

\section{Fluctuation Theorem} 
The  spectral budget in Eq. \eqref{eq:budg} can be interpreted as a nonlinear Fokker-Planck equation (FPE) with scalewise separated source/sink terms. Accordingly, the underlying cascade can be expressed as a stochastic process \citep{crispin1994handbook}, whereby trajectories represent time histories of energy packets (eddies) traveling in $k$-space driven by advection and diffusion, until they disappear by virtue of a killing term linked to the action of viscous dissipation. The latter term absorbs trajectories as a state-dependent Poisson process with a rate $2\nu k^2$ \cite{daly2007intertime}. 

If the steady-state solution is known (Eq. \ref{eq:Sol_emp}), the corresponding drift and diffusion for the position in $k$ space of the energy packet can be formulated as a function of $k$. Thus, a Langevin equation that ensures that the steady-state probability density function (PDF) abides by Eq. \eqref{eq:Sol_emp} is 
\begin{multline}
dk= k\left(4\tau_* - k \frac{d\tau_*}{dk}\right) \tau_*^{-2}  dt+ b(k) dW,
\label{eq:stoch_Leith}
\end{multline}
where $dW$ is the Wiener increment and $b(k)=\sqrt{2} k \tau_*^{-\frac{1}{2}}$.  This equation is subjected to a unit rate of birth at $k=k_i$ and a state dependent killing term with rate $2\nu k^2$ \citep{daly2007intertime}.  Since the FPE is written in the so-called transport form, the interpretation of the multiplicative term is the one of Hanggi-Klimontovich \citep{hanggi1978stochastic,klimontovich1990ito,porporato2011local}.  

\begin{figure}
\begin{centering}
\includegraphics[scale=0.5]{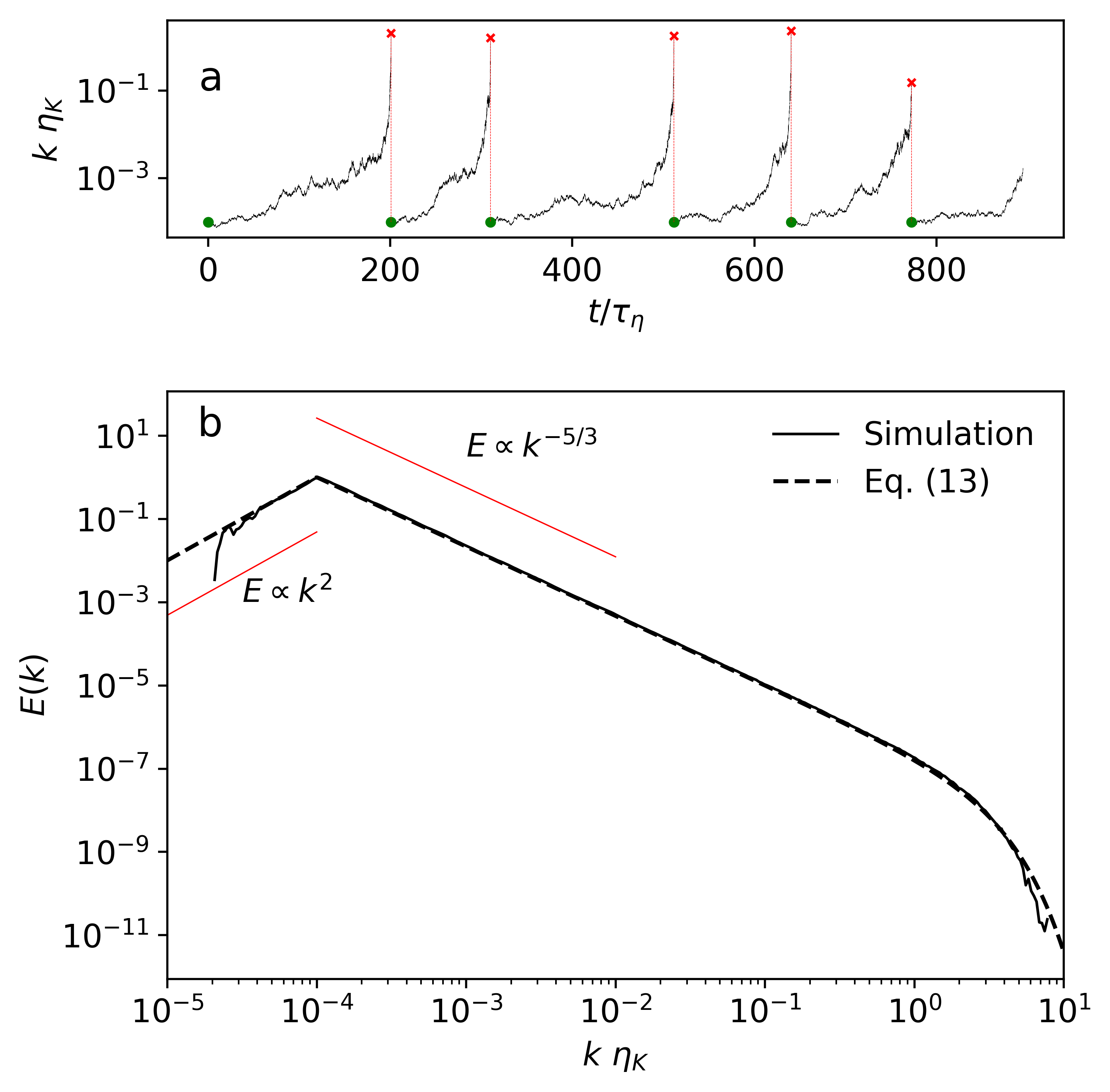}
\caption{(a) A numerical realization of the stochastic process given by Eq. \eqref{eq:stoch_Leith}. The trajectories are terminated following a state-dependent Poisson process with rate $2\nu k^2$ and initiated at $k_i \eta_K = 10^{-4}$. (b) The steady-state PDF from the numerical simulation and the approximate spectrum in Eq. \eqref{eq:Sol_emp} are compared. For reference, the red lines show $k^{-{5}/{3}}$ (K41 or cold cascade) and $k^{2}$ (warm cascade or Saffman spectrum).}
\label{fig:spectrum}
\end{centering}
\end{figure}

With this formal correspondence, the statistics of irreversiblity of Eq. (\ref{eq:stoch_Leith}) can now be analyzed \citep{porporato2007irreversibility, porporato2011local} for steady-state homogeneous and isotropic turbulence with energy injected at $k_i$ and transported on average towards higher wavenumbers where dissipation takes place. This allows an illustration of the fluctuation theorem for fully developed turbulence fluctuations, linking the turbulent entropy balance at $k$ to those of forward and backward energy cascades.  Fig \ref{fig:spectrum}a shows a numerical realization of this process in which energy packets are injected at $k_i\eta_K= 10^{-4}$ after the termination of trajectories by dissipation. While the energy injection at $k_i$ is only related statistically to the dissipation, the immediate re-injection after killing adopted here for convenience of simulation and visualization preserves the steady state PDF. As shown in Fig \ref{fig:spectrum}b, the steady state PDF of the $k$ time series corresponds to the empirical spectrum in Eq. \eqref{eq:Sol_emp}. 

The injection of energy at lower $k$ and the dissipation sink at higher $k$ produce a non-zero average current and a non-equilibrium steady state (NESS) current of energy towards smaller scales. The stochastic fluctuating velocity $\dot{k}|k$ is a random variable with mean current velocity given by \citep{porporato2011local} 
\begin{equation}\label{eq:v_general}
v_{NESS}(k)= \frac{J(k)}{E(k)} \approx 2k\tau_*^{-1}-k^2E_*^{-1}\tau_*^{-1}\frac{dE_*}{dk}, 
\end{equation}
and a fundamental FT-type symmetry \citep{porporato2007irreversibility}.
The degree of irreversibility of the NESS resulting from the cascade towards dissipation may be given by the rate of energy transfer to smaller scales. According to the formalism of stochastic thermodynamics \citep{seifert2012stochastic, jarzynski2011equalities}, the non-zero current velocity may be associated with a positive `turbulent entropy' production rate $\Sigma$ \citep{porporato2011local},  
\begin{equation}\label{eq:s_general}
\Sigma(k)=2\left[\frac{v_{NESS}(k)}{b(k)}\right]^2  \approx \left( \frac{dE_*}{dk} - 2E_*\right)^2  E_*^{-2} \tau_*^{-1}.
\end{equation}

In the inertial sub-range where $E(k)= C_o \epsilon^{2/3} k^{-5/3}$ (i.e. $f_{\eta}(k\eta_K)\approx 1$), $J=(11/3) C_o^{3/2} \epsilon$ (a constant), $v_{NESS} =\sqrt{C_o}\epsilon^{1/3} k^{5/3}$ and $\Sigma=(121/9) \sqrt{C_o}\epsilon^{1/3} k^{2/3}$. These scaling laws have also been confirmed by simulations (not shown) of the stochastic process in Eq. \eqref{eq:stoch_Leith}, Eqs. \eqref{eq:v_general} and \eqref{eq:s_general} using the model spectrum and relaxation time scale in Eqs. \eqref{eq:Sol_emp} and \eqref{eq:emp_tau}. 

\section{An Extended Turbulence Thermodynamics}

The term $\Sigma(k)\ge 0$ can be shown to be the source term in the balance equation for the turbulent entropy $S_t=-\ln E$  \cite{seifert2005entropy}. In this view, the asymmetry in the turbulence cascade that not only transports energy forward to higher wave numbers (the forward turbulence cascade) but also backward (the so-called back scatter) contributing energy to larger eddies, is also linked to a turbulent entropy production. Specifically, if one considers the (random) path of a turbulent energy packet over a period of time $[0,t]$, the ensemble average of the ratio of the path measures between forward and backward cascade, that is the path measures $P_f, P_b$ the forward and backward PDFs of $k(t')$ evaluated along stochastic trajectories that take place on a common interval $[0,t]$ in steady state conditions \cite{maes2003time, porporato2011local}, is
\begin{equation}
\Bigg\langle \ln \frac{P_f}{P_b}\Bigg\rangle= \int_0^t\int_0^\infty  E(k,t')\Sigma(k,t')dkdt'.   \end{equation}
Thus, $S_t$ can be interpreted as a scale-wise `turbulence entropy', which together with TKE may be used to define an extended turbulence thermodynamics as in Richardson's  \citep{richardson1920supply}, providing a measure of the number of turbulent states at wave number $k$ to be linked to the corresponding portion of TKE. Neglecting momentarily the viscous subrange, TKE can be linked to the area under the spectrum given in Eq. \eqref{eq:sol_leith_sim}, namely

\begin{equation}\label{eq:K}
\begin{split}
K = &  2 \int_0^{1/\eta_K} E dk = \\
& 3 C_o  \nu^2 k_i^2 Re^2  \left(\frac{11}{9} - {Re}^{-1/2}\right),
\end{split}
\end{equation}
where $Re=l_i u/\nu$ is a Reynolds number formed from a characteristic length $l_i=1/k_i$ and velocity $u=(\epsilon/k_i)^{1/3}$. This definition ensures that $l_i/\eta_K={Re}^{3/4}$ consistent with expectations for many turbulent flows \citep{tennekes1972first}.  The integrated  entropy $S_I=\int S_t(k)dk$ can also be obtained as 

\begin{equation}
\label{eq:SI}
\begin{split}
 S_I =  &-\int_0^{1/\eta_K} \ln{E}dk = \\
 & \frac{11}{3} k_i -  k_i {Re}^{3/4} \left[\frac{5}{3} + \ln \left( C_o \nu^2 k_i {Re}^{3/4} \right)  \right].
\end{split}
\end{equation} 

One objection that can be raised here is that $S_t$ is not an entropy in the sense of Clausius, but an extended entropy related to macroscopic turbulent fluctuations. However, similar to classical thermodynamics, we may consider the TKE to correspond to an internal kinetic energy of turbulence, so that ${dS_I}/{dK}={1}/{T_I}$, analogous to an effective temperature for the turbulent system.  When such a definition is combined with Eq \ref{eq:K}, the extended turbulence thermodynamics can proceed as follows: 

\begin{equation}
\label{eq:SI}
\begin{split}
T_I = & \frac{dK}{dRe} \left(\frac{dS_I}{dRe}\right)^{-1} =   \\
& 18 C_o \nu^2 k_i \frac{ {Re}^{5/4} \left(  {Re}^{-1/2} - \frac{44}{27}\right)}{8+ 3 \ln \left(C_o \nu^2 k_i {Re}^{3/4}\right)}.
\end{split}
\end{equation} 
 
\begin{figure}
\begin{centering}
\includegraphics[scale=0.5]{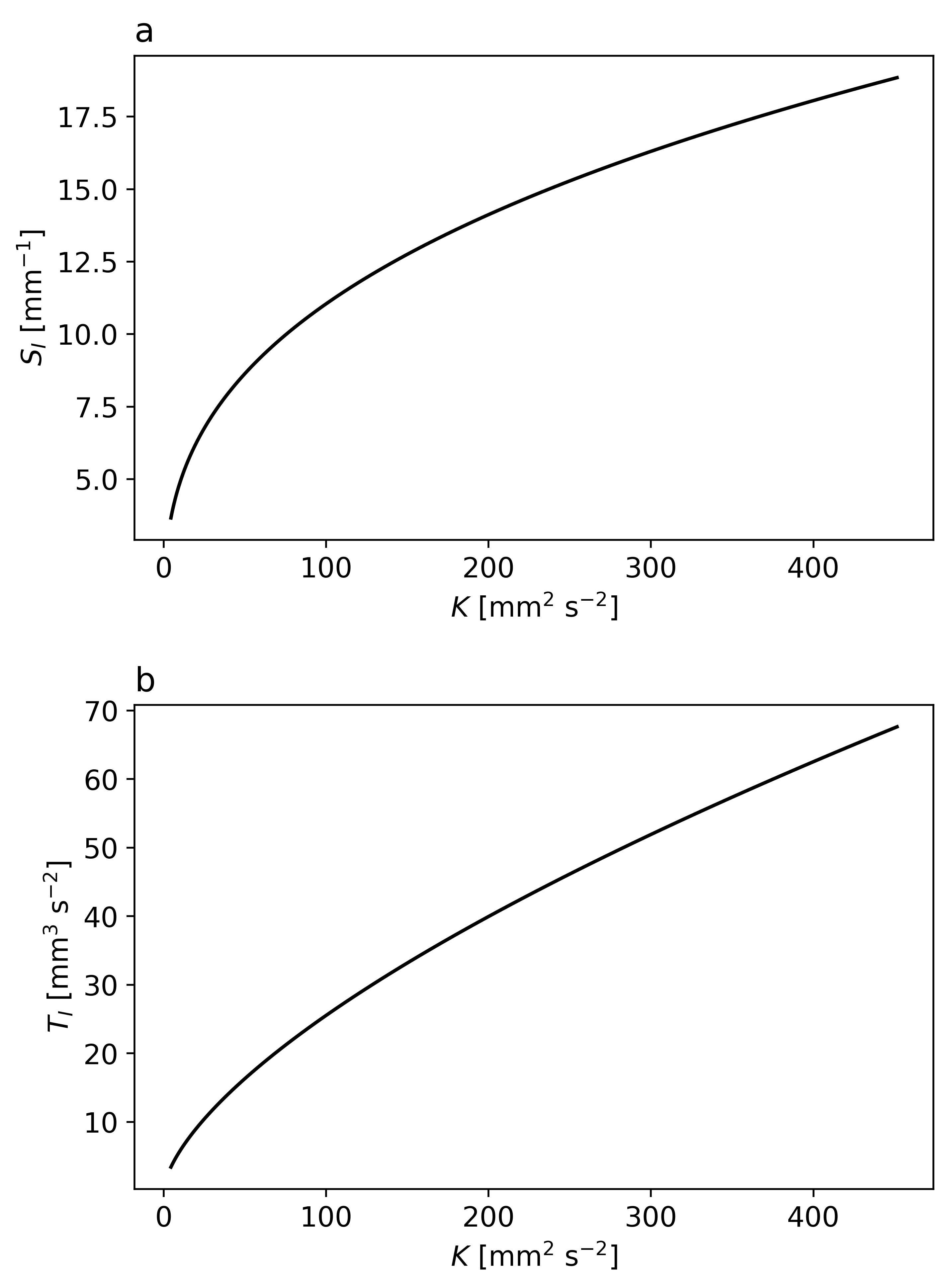}
\caption{The relation between entropy $S_I$ (a) and turbulent temperature $T_I$ (b) with TKE $K$. For these results, it was assumed that $k_i=1\ \text{m}$ and $\nu= 8.95 \times 10^{-7}  \text{m}^2/ \text{s}$. }
\label{fig:entropy}
\end{centering}
\end{figure}

An integrated positive turbulence-entropy production for the integrated turbulence entropy, $S_I$, can be readily obtained assuming that molecular dissipation primarily acts in the neighborhood of $k=1/\eta_K$. For  $k<k_i$ where $E \approx C_2  k^{2}$ and $f_{\eta}(k\eta_K)\approx 1$, $v_{NESS} \approx 0$ and $\Sigma \approx 0$. As a result, 

\begin{equation} 
\label{eq:Sigma_I}
\begin{split}
\Sigma_I &  = \int_{0}^{1/\eta_K} E\ \Sigma dk  = \\
& \frac{121}{12}  C_o^{3/2} \nu^3 k_i^4 {Re}^{3} \ln{Re}.
\end{split}
\end{equation}
In the limit of high $Re$ the above relationships approach to  $K \propto Re^{2}$, $\Sigma_I \propto  Re^{3}$, $S_I \propto Re^{3/4-\alpha_1}$, and  $T_I \propto Re^{5/4+\alpha_2}$, where $\alpha_1$ and $\alpha_2$ are deviations due to the logarithmic terms. At Re $\approx 1$, the spectrum within the inertial subrange vanishes with production scales being commensurate to the Kolmogorov microscale ($\eta_K k_i \approx 1$). For this case, $\Sigma_I \approx 0$ ({\bf detailed balance holds}) although the cascade still generates small $K$, $S_I$, and $T_I$. The key quantities in this turbulent thermodynamics are plotted in Fig. \ref{fig:entropy}. The fundamental equation, $S_I=S_I(K)$ shows a downward concavity that ensures entropy production by `combining' turbulent flows of different TKE, while the dependence of $T_I$ on $K$ shows a turbulent TKE capacity ($dT_I/dK$) that is not constant but decays with the Reynolds number.

\section{Conclusions} 

The fluctuation theorem has been used to analytically link the shape of the energy spectrum with the imbalance between forward and backward probabilities of energy packets moving scale-wise in time across the energy cascade.  The difference between these two aforementioned probabilities is the main cause why the `detailed balance' or `microscopic reversibility' (i.e. at equilibrium, each elementary process is in equilibrium with its reverse process) is not applicable to turbulence, and why turbulent fluctuations are presumed to be far from equilibrium. 

The previous result unfolded a connection between the turbulent entropy production rate measuring the effective spreading of energy-packet trajectories in the cascade, the thermodynamic entropy production, and the Reynolds number $Re$ for an externally prescribed injection scale $1/k_i$ (often dictated by boundary conditions or geometry).  As first pointed out by Landau \citep{landau1959fluid} and substantiated in later studies \citep{constantin1985determining}, the finite $Re$ is an indicator of the number of degrees of freedom of the turbulence cascade, $N_d\sim (l_i/\eta_K)^{3}\sim {Re}^{9/4}$.

An additional foresight from this analysis is that the shape of the spectrum at low $k$. It is shown here that the Saffman spectrum is linked to $v_{NESS}=0$ and $\Sigma=0$ (no scale-wise entropy production and the detailed balance is satisfied as expected for warm cascades). From $J$ in Eq. \eqref{eq:cont}, the condition for $v=J/E \neq 0$ assuming $E(k) \propto k^{\gamma}$ can now be derived for the large scales ($k<k_i$). With $\gamma>0$, the condition $-dJ/dk >0$ imposed by the energy balance necessitates $\gamma > \alpha$ (=2 for the Saffman spectrum and the associated Leith's model) for $k/k_i<1$. It also follows that $J<0$ (or $v_{NESS} <0$) when $\gamma<\alpha$, a state where the current towards larger scales is caused by the dominance of the backscatter over the forward drift.  From the perspective of $\alpha=2$ (i.e. Leith's model), the Saffman ($\gamma=2$) spectrum results in $-dJ/dk=0$ whereas the Batchelor \citep{batchelor1956large} spectrum ($\gamma=4$) yields $-dJ/dk>0$ (i.e. forward drift still dominates over backscattering).  However, the Karman spectrum \citep{meyers2008functional,pope2001turbulent} often used in reshaping the inertial subrange spectrum at production scales in boundary-layer turbulence yields a non-monotonic $-dJ/dk$ in the rising limb of $E(k)$ as $k\rightarrow k_i$.    

These considerations have also led to a new perspective on the turbulence-thermodynamics formalism linking the emergence of turbulent modes to store disorderly kinetic energy to key macroscopic quantifies such as the Reynolds number and the turbulence temperature. It will be of interest to compute $S_I$ and $T_I$ based on an energy spectrum including intermittency corrections, to assess how intermittency might play a role in the proposed extended thermodynamics.  One might also conjecture the existence of an extended global turbulence pressure to link flow configurations to kinetic energy and entering as a natural variable in a Gibbs turbulent free energy to provide a unified criterion for turbulent transition and development. It is hoped that future investigations will contribute further elements to picture of turbulence as a nonequilibrium phase transition, of which several elements are beginning to emerge \citep{goldenfeld2017turbulence}. 

\begin{acknowledgments}
A.P. acknowledges support from the US National Science Foundation (NSF) grants EAR-1331846 and EAR-1338694 and the Carbon Mitigation Initiative at Princeton University. M.H acknowledges support from the Princeton Institute for International and Regional Studies and the Princeton Environmental Institute. G.K. acknowledges support from the NSF grants AGS-1644382 and IOS-1754893.
\end{acknowledgments}

\bibliography{canopy2D}
\bibliographystyle{apsrev4-1}

\end{document}